\documentclass[pdflatex,sn-basic]{sn-jnl}

\usepackage{graphicx}%
\usepackage{multirow}%
\usepackage{amsmath,amssymb,amsfonts}%
\usepackage{amsthm}%
\usepackage{mathrsfs}%
\usepackage[title]{appendix}%
\usepackage{xcolor}%
\usepackage{textcomp}%
\usepackage{manyfoot}%
\usepackage{booktabs}%
\usepackage{algorithm}%
\usepackage{algorithmicx}%
\usepackage{algpseudocode}%
\usepackage{listings}%

\begin{document}

\title[FIEVel: a Fast InExpensive Velocimeter based on an optical mouse sensor]{FIEVel: A Fast InExpensive Velocimeter based on an optical mouse sensor}

\author*[1]{\fnm{Robert} \sur{Hunt}}\email{robert\_hunt@brown.edu}

\author[1]{\fnm{Eli} \sur{Silver}}\email{eli\_silver@brown.edu}

\author*[1]{\fnm{Daniel M.} \sur{Harris}}\email{daniel\_harris3@brown.edu}

\affil[1]{\orgdiv{School of Engineering}, \orgname{Brown University}, \orgaddress{\street{184 Hope St.}, \city{Providence}, \postcode{02912}, \state{RI}, \country{USA}}}

\abstract{Fluid velocimetry is fundamental to a breadth of applications spanning academia and industry, however velocimetry at high temporal resolution is often prohibitively costly. Here, we introduce a Fast and InExpensive Velocimeter (FIEVel) based on an optical mouse sensor. At its core, the optical mouse sensor consists of a small pixel array that acquires image data at high rates, with onboard hardware to compute and output motion in two orthogonal axes directly. By adapting this widely available integrated circuit to fluid velocimetry, we demonstrate that FIEVel is capable of resolving two components of velocity non-intrusively in the bulk of a flowing fluid at rates up to 6.4 kHz, at orders of magnitude lower cost than traditional velocimetry devices. We demonstrate and validate the velocimeter across a range of flows, operating conditions, and illumination strategies. As a sample application, we measure the temporal spectrum of a grid-generated turbulent flow and show excellent agreement with traditional Particle Image Velocimetry (PIV) using a high-speed camera.}

\keywords{fluid velocimetry, particle image velocimetry, optical methods}

\maketitle

\section{Introduction}\label{sec:Intro}
\begin{figure*}[h]
\includegraphics[width=\textwidth]{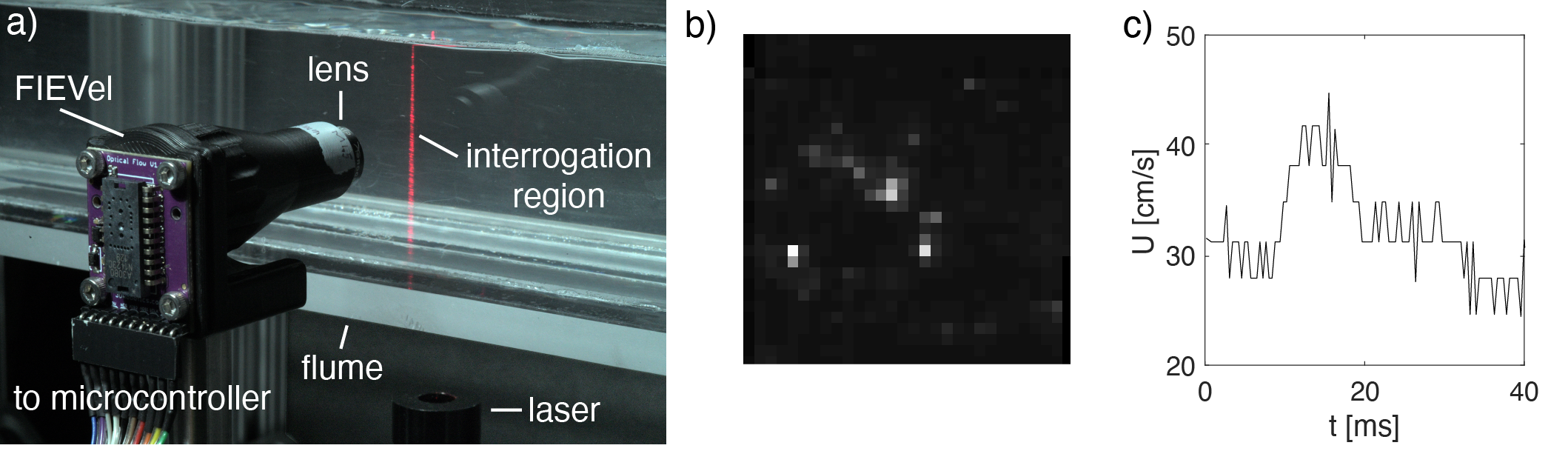}
\caption{(a) Schematic of FIEVel and flume experiment used for characterizing the device. A high-speed camera is positioned across the channel from FIEVel, and both devices image a region illuminated from below by a constant-output laser. (b) Sample image output from FIEVel.  (c) Example time series of measured horizontal velocity, collected at 2.5 kHz sample rate with 0.8 mm field of view and 63 $\mu$s exposure setting.}\label{fig:overview}
\end{figure*}

The flow of a fluid is characterized by its local velocity. In the field of fluid mechanics, the ability to measure such velocities (i.e. velocimetry) is fundamental to nearly every experimental effort.  In response to this widespread need, a range of commercial and bespoke solutions have been developed targeted to different applications.

Traditional systems for high-speed velocimetry typically cost over $\$10,000$, with many exceeding $\$100,000$, and each method has its own limitations that can be prohibitive for the required measurement (\cite{ExperimentHandbook}). Standard techniques like hot wire anemometry, laser doppler velocimetry, and acoustic doppler velocimetry can be limited by intrusiveness, sample rate, signal quality, and/or their ability to resolve multiple components of velocity. Although some lower cost methods for velocimetry exist, such as rotating vane anemometers and pitot tubes, these are often have strong limitations related to transient response, measurement volume, and sensitivity.

One of the most common and widely accepted techniques for measuring fluid velocities in practice is Particle Image Velocimetry (PIV). Conceptually, images of a flowing, particle-seeded fluid are compared using a cross-correlation algorithm from which flow velocities can be computed from measured displacements. The sample rate of this measurement is limited by the camera frame rate. While some low-cost PIV solutions have emerged more recently (\cite{ryerson2012simple,ring2013development,GoProPIV}), the current readily available solutions that can achieve fast time resolution ($\gtrsim$ 1 kHz, essential for many fundamental scientific studies and applications) are extremely expensive. High-speed cameras used to achieve high temporal resolution (\cite{CallumGray, PostageStamp}) are typically tens of thousands of dollars. Further, these captured images are typically stored locally in RAM on the camera and downloaded to a PC for processing, limiting the amount of data that can be acquired and imposing an additional cost. Such technologies are largely inaccessible to many labs, institutions, and technical applications, potentially limiting progress in the field.

Coincidentally, an optical computer mouse operates with a similar working principle to PIV, using a low-resolution image sensor and cross-correlation algorithm to compute displacements between frames at very high rates (typically $\gtrsim$3 kHz for a standard mouse, priced $<$\$10).  Although a typical PIV system can measure the velocity field at multiple locations over a large field of view, the mouse sensor measures pointwise x-y velocities at a much higher rate for many orders of magnitude lesser in cost. Further, the velocity calculation is performed onboard the sensor, significantly lowering the required bandwidth compared to streaming image data to a computer or RAM. Although optical mouse sensors have been adapted generally for 2D-displacement measurements (reviewed in \cite{displacement}), they have not been applied to measure velocities of a fluid to the best of our knowledge.

In the current work, we adapt a commercially available optical mouse sensor to perform velocimetry in the bulk of a flowing fluid by developing a custom imaging, illumination, and processing system, with total cost around $\$100$. We demonstrate the capabilities of our Fast, InExpensive Velocimeter (FIEVel) by measuring a range of flows under varying operating conditions and comparing with a state-of-the-art PIV system that uses a high-speed camera. As an example application, we measure the temporal spectrum of a grid-generated turbulent flow.

\section{Methods}\label{sec:Methods}

\subsection{FIEVel}
FIEVel uses an ADNS3080 optical mouse sensor installed on a breakout board (EC Buying ADNS-3080). This sensor has 30 x 30 px resolution and provides displacement data with 0.5 px precision.  A custom 3D printed housing and removable lens system, printed using a Bambu Lab X1-Carbon FDM printer, is designed for variable magnification and ease of use. A 25 mm focal-length lens (Wendry 25 mm) is installed at various distances from the image sensor to adjust magnification, allowing for a working distance of up to 50 mm for the configurations considered here. To avoid vignetting effects, a physical aperture on the ADNS3080 IC is removed. The device is pictured in Figure \ref{fig:overview} (a), with typical image in (b) and calibrated displacement output in (c). The ADNS3080 is connected over SPI to a PJRC Teensy 4.1, and the Teensy sends data to a computer through a USB serial connection. The $LED\_CTRL$ output from the ADNS3080, used for illumination control of the LED in a typical optical mouse, is connected to the Teensy in order to trigger an interrupt routine that reads motion data from the sensor and outputs the data over serial, allowing for frame-synchronized data acquisition. The sensor, lens, microcontroller, and laser are available for just over $\$100$ in total at the time of writing. Details needed to replicate the device can be found at \href{https://github.com/harrislab-brown/FIEVel}{https://github.com/harrislab-brown/FIEVel}. 

FIEVel's data acquisition software is adapted from from the RCMags ADNS3080 library on GitHub (\cite{RCMags}). Acquisition is performed with Automatic Gain Control (AGC) off, Fixed Frame Rate control on, LED shutter mode on, and resolution set to high. Image acquisition, used for alignment, focusing, and calibration, is performed utilizing $Pixel\_Burst$ mode, and displacement data acquisition is performed using $Motion\_Burst$ mode. To ensure regular and synchronized displacement data acquisition, motion burst requests are embedded in an interrupt service routine that is triggered by the $LED\_CTRL$ output. Acquisition times are recorded from the Teensy internal clock.

\subsection{Validation}
Experiments are performed in the OpenFlume tabletop, open-channel water flume with a 5.0 cm wide channel detailed and characterized in prior work (\cite{hunt2023,Lewis2024}). A commercially available red laser (oxlasers 767461754339) with a continuous output is used with its inline cylindrical lens removed. The laser is aligned vertically in the center of the test section channel, 17.0 cm downstream from the beginning of the test section, and focused to a spot diameter of around 2 mm. A high-speed camera (Phantom VEO 440L) equipped with a Nikon AF Micro-NIKKOR 200 mm lens, at 1x magnification, is used to capture images at 5000 fps with exposure 141 $\mu$s, unless stated otherwise. FIEVel is installed directly across the channel from the Phantom camera, and both are focused on the same measurement region located 30 mm above the channel bottom as identified by a calibration slide target. Potters Industries Sphericel 110P8 hollow glass microspheres with mean density 1.10 g/cm$^{3}$ and median diameter $10$ $\mu$m are used as tracer particles.

Both FIEVel and the high-speed camera are calibrated by concurrently imaging a standard microscope calibration slide with a circular target. To ensure uniformity between imaging modalities, each image was intensity corrected so that the center of the calibration target circle appears black and the background appears white. This corrected image of a circle is fit using the MATLAB \textit{imfindcircles} routine.  

Images captured from the Phantom high-speed camera were processed in PIVLab (\cite{PIVLab}). Images were preprocessed using Contrast Limited Adaptive Histogram Equalization (CLAHE) with a window size of 64 px and automatic contrast stretching enabled. A 64 x 64 px interrogation window size is used with multiple passes of the same size repeated until quality slope reaches 0.025, with 2 x 3-point gaussian sub-pixel estimator and standard correlation robustness settings.

To generate turbulence in the OpenFlume channel, a 3.0 mm thick acrylic sheet was laser cut in a square grid pattern with 4.0 mm x 4.0 mm square openings repeated in 5.0 mm intervals (with 1.0 mm solid material in between). This grid, 8.0 cm x 5.0 cm overall in size, was installed at the entrance of the test section, 17.0 cm upstream from the measurement location.

In addition to illuminating the fluid and tracer particles with a laser oriented orthogonally to the camera view, two other illumination schemes were used for velocimetry with FIEVel. First, a laser oriented obliquely to the optical axis (at approximately 45 degrees) is used, shining through the same acrylic panel which the optical measurements are performed. Second, a 5W white LED is collimated and oriented directly towards FIEVel along the optical axis in a typical particle shadow velocimetry (PSV) configuration.

\section{Results}\label{sec:Results}
\subsection{Flow measurements}

\begin{figure*}[h]
\centering
\includegraphics[width=\textwidth]{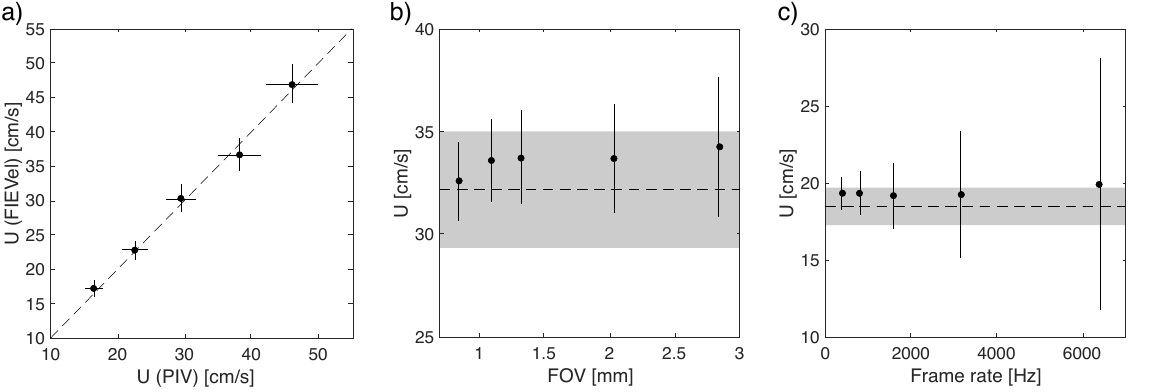}
\caption{(a) Measured velocity and standard deviation as a function of flow rate for fixed field of view (magnification) and frame rate. The horizontal axis represents the benchmark velocity as measured from high-speed PIV, and the vertical axis represents the velocity measured by FIEVel. The diagonal is represented by a dashed black line. FIEVel measurements were recorded at 1.5 kHz with 1.1 mm field of view and 63 $\mu$s exposure.  (b) Measured velocity and standard deviation as field of view is varied for a fixed flow speed and frame rate. Mean value from PIV is indicated by a dashed black line with standard deviation shaded gray. FIEVel data are acquired at 1.0 kHz, with exposure time set to 21 $\mu$s for the largest two fields of view and 63 $\mu$s for the smallest three.  (c) Measured velocity and standard deviation as a function of frame rate for a fixed flow speed and field of view. Mean value from PIV is indicated by a dashed black line with standard deviation shaded gray. FIEVel data was collected at 1.3 mm field of view and 42 $\mu$s exposure.}\label{fig:Sweeps}
\end{figure*}

To assess the capability of FIEVel, we performed direct velocimetry measurements under typical operating conditions of the OpenFlume facility (without grid turbulence), characterized in prior work (\cite{hunt2023,Lewis2024}) as largely uniform across the measurement region and with low turbulence intensity. We compared the measured velocity output from FIEVel with results from standard PIV measurements that utilize a high-speed camera (described in Methods) imaging the same location in the flume and illuminated by the same source. We performed this validation procedure across several flow conditions, magnifications, and frame rates, varying one parameter while keeping others fixed. The result of these experiments is summarized in Figure \ref{fig:Sweeps}.

In Figure \ref{fig:Sweeps}(a), the flow rate was varied from 17 - 46 cm/s and measured for each case by both FIEVel and high-speed PIV while keeping the field of view and frame rate constant. Here, the mean velocity measured by FIEVel shows good agreement with PIV, with a mean absolute error of 2.9\% and a maximum absolute error of 4.9\%. Here, the flume turbulence intensity and subsequent measured standard deviation increases as the flow speed increases, as indicated by the benchmark measurements and corroborated by FIEVel measurements.

In Figure \ref{fig:Sweeps}(b), the field of view of FIEVel is varied over a range of 0.8 to 2.8 mm by adjusting the lens to image sensor distance (effectively changing the magnification) while keeping the flow speed and frame rate fixed. FIEVel velocity measurements of the mean show good agreement with high-speed PIV, with an average absolute error of 4.5\% and a maximum of 6.6\%. Since the flow speed and frame rate are fixed, decreasing field of view results in a larger displacement in pixels reported by FIEVel, effectively increasing the magnitude of the signal relative to the quantization increment.

In Figure \ref{fig:Sweeps}(c), the frame rate of FIEVel is varied from 400 Hz to 6.4 kHz for a fixed flow speed and field of view. The mean velocity shows good agreement, with an average absolute error of 5.0\% and with the worst case at 7.8\%. In order to span the entire range of frame rates available to the sensor while keeping the flow speed and field of view fixed, the highest frame rate case was restricted to around 0.7 pixels displacement per frame, yet the measured mean value is surprisingly robust in spite of the comparable 0.5 pixel output quantization increment. However, the standard deviation reported from FIEVel continues to increase as the mean displacement approaches the quantization size. The error due to this effect is discussed in the Discussion section. 

\begin{figure}[h]
\centering
\includegraphics[width=.7\textwidth]{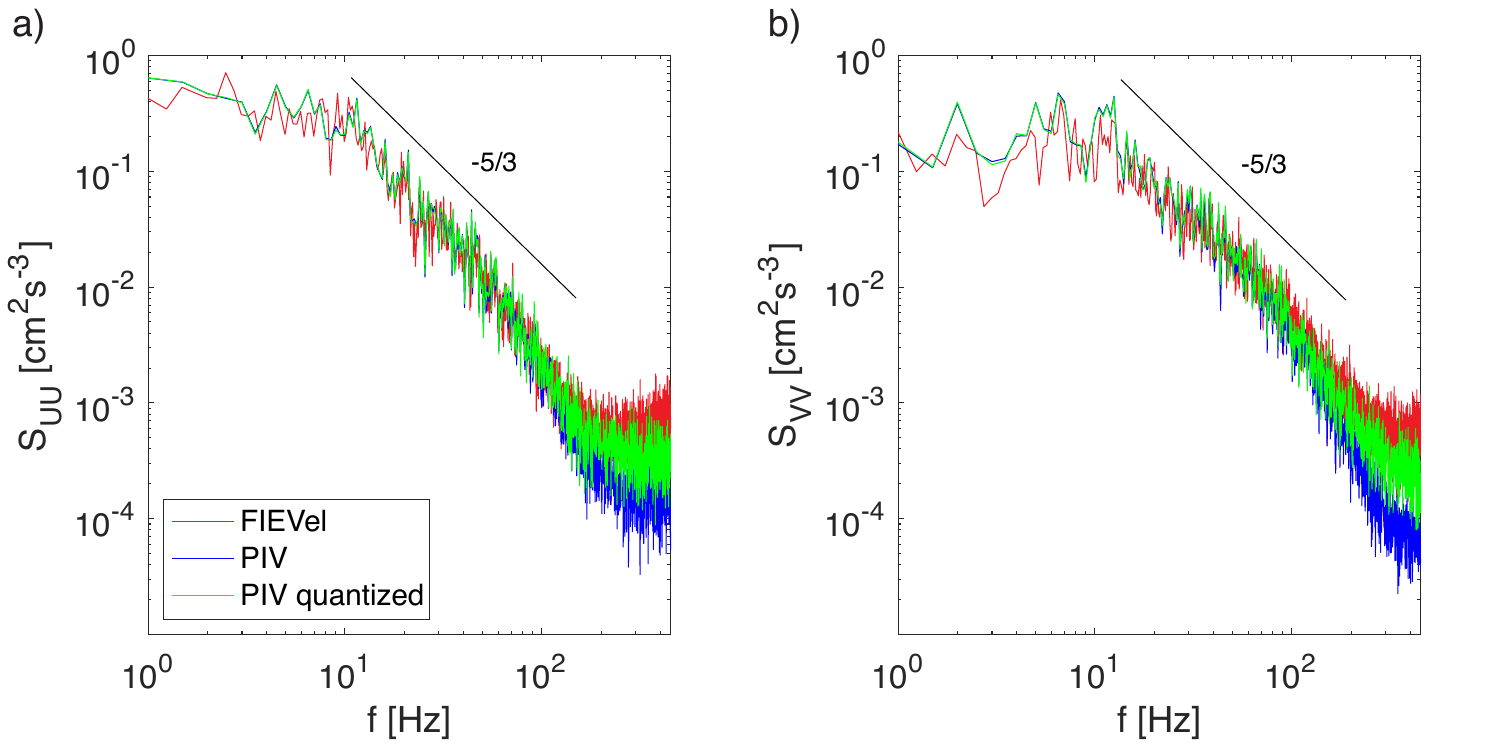}
\caption{Power spectral density $S_{u_{i} u_{i}} = \frac{|fft(u_{i})^2|}{N f_{s}}$ for FIEVel (red), PIV (blue), and artificially quantized PIV (green) as a function of frequency. Horizontal velocity $U$, presented in (a), has mean 29.5 cm/s and standard deviation 3.85 cm/s. Vertical velocity $V$, presented in (b), has mean 1.15 cm/s and standard deviation 3.49 cm/s. FIEVel data is collected at 2.5 kHz sample rate with 0.8 mm field of view (FOV) and 63 $\mu$s exposure setting.}\label{fig:Turb}
\end{figure}

As an example application, velocity measurements are performed in a grid-generated turbulent flow, with experimental details provided in the Methods section. In Figure \ref{fig:Turb}, we plot the power spectral density $S_{u_{i} u_{i}} = \frac{|fft(u_{i})^2|}{N f_{s}}$, where $u_{i}$ is a fluctuation from the mean velocity in the $i$-th component, $N$ is the number of discrete samples, and $f_{s}$ is the sample frequency. We consider each signal over nine intervals of $N=10,000$ discrete measurements with 50\% overlap and average these to yield the curves plotted in Figure \ref{fig:Turb}. A subset of the FIEVel data is presented as a time series in Figure \ref{fig:overview} (c).

To assess the possible influence of quantization error, we artificially quantized the velocity data captured from the high-speed camera to match the quantization size inherent to FIEVel measurements. The output from FIEVel, high-speed PIV, and artificially quantized high-speed PIV all show good agreement with each other, and they resolve the predicted scaling for isotropic turbulence in the inertial subrange (\cite{pope2000turbulent}) as indicated by the -5/3 power law reference curve in Figure \ref{fig:Turb}. Somewhat surprisingly, even for much more coarse quantization than shown here, the frequency spectrum is relatively robust compared to that of the full resolution output. This demonstrates promise for the use of FIEVel in turbulence spectrum measurements where cost has been a primary limiting factor.

As in PIV, the frequency resolution of the sensor is limited by spatial filtering due to size of the measurement region relative to spatial fluctuations in the velocity field (\cite{SpatialFilter}). This limit can be estimated as $U/L$, where $U$ is the mean flow and $L$ is the field of view. The spectrum is plotted up to this value.

\begin{figure}[h]
\centering
\includegraphics[width=0.4\textwidth]{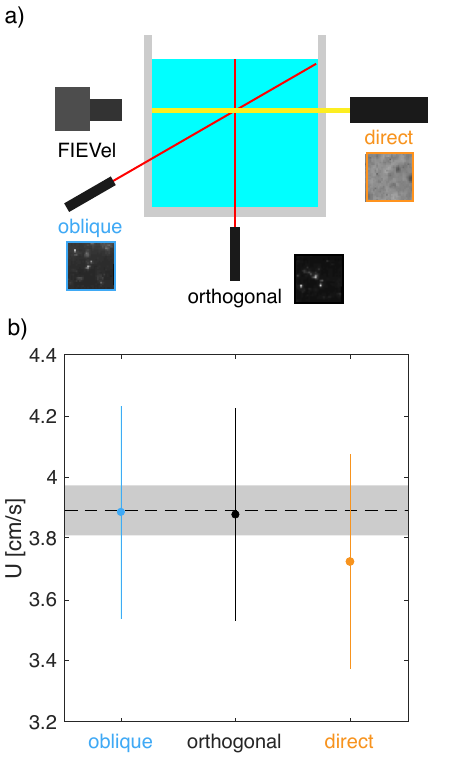}
\caption{(a) Measurement configuration for three illumination schemes, with FIEVel facing the flume from the side. In prior data, the orthogonal laser illumination configuration was used (black).  The oblique configuration corresponds to laser illumination at 45 degrees from the optical axis (blue). For direct illumination (orange), a collimated white LED is aimed directly towards FIEVel from across the flume test section. Panel (b) depicts the measured velocities observed for each illumination scheme depicted above under the same flow conditions. All FIEVel data is collected at 0.5 kHz and 0.8 mm field of view. Oblique and orthogonal data is collected at 10 $\mu$s exposure setting, direct is collected at 1.0 $\mu$s exposure setting. PIV benchmark data is acquired at 0.5 kHz.}
\label{fig:methods}
\end{figure}

Furthermore, FIEVel is able to measure velocity with a variety of illumination schemes, in contrast with traditional PIV where a thin laser sheet is aligned orthogonally to the optical axis. The data presented thus far uses the orthogonal laser illumination scheme. An oblique (45 degree) laser illumination scheme is also tested, wherein the laser is shone through the same acrylic panel from which imaging is performed. A direct illumination scheme, comparable to particle shadow velocimetry (PSV), is depicted in the upper panel of \ref{fig:methods}, where the light from a collimated white LED is shone directly towards the image sensor. Although PSV images are typically inverted before processing, FIEVel is capable of velocity measurements under this illumination condition without inversion. The mean flow speed from all three illumination strategies is in good agreement with the PIV benchmark.

\section{Discussion}
Fluid velocimetry is essential for research, industry, and education, yet high-speed non-intrusive measurements are unattainable for many applications. FIEVel makes these measurements possible by applying the established principles of PIV to a mass-produced optical displacement sensor. We demonstrate the capabilities of FIEVel for pointwise velocity measurements in the bulk of a flowing fluid across flow conditions and sensor operating conditions, including variations in flow speed, optical magnification, and frame rate.  Our device is validated via direct comparison to traditional PIV measurements with a high-speed camera under identical illumination and flow conditions. Further, we demonstrate the capability of FIEVel to acquire measurements under various illumination strategies, beyond those that are typically applied in traditional PIV. As the field of view of FIEVel is usually equivalent to or smaller than a typical laser spot size, this removes the need for sheet illumination with high-power lasers as well as the subsequent angular alignment and focusing, which can increase the difficulty in implementation, risk to users, power requirements, and overall cost. In addition to established techniques of orthogonal (PIV) and direct (PSV) illumination, we demonstrate FIEVel's ability to utilize oblique lighting originating near the sensor, generally improving access and versatility in application.

Although the range of operating conditions presented here are broad, these are certainly not exhaustive, with many limitations imposed by the selection of a specific sensor, optical configuration, illumination device, and test facility. Some limitations and possible improvements to the device are discussed below.

\subsection{Limitations and future work}
As the image sensor is relatively low resolution, there is an inherent limit on the resolution of the displacement output. Here, the image sensor is 30 x 30 pixels, and the sensor uses subpixel estimation to yield an effective displacement output resolution of 0.5 pixels. Thus, the theoretical output range of the sensor is nearly 7 bits, while in practice the output range is 6 bits or less. For a given flow condition, the user should tune the field of view and frame rate to optimize displacement between frames, thereby utilizing the full range of the sensor and minimizing quantization error.

Quantization error is caused by the discrete approximation of a continuously varying input, and here this error strongly imposed by rounding the observed displacement between frames to the nearest half pixel. For a velocity distribution whose variance is large relative to the quantization size, the mean squared error induced by quantization can be estimated as $\frac{\Delta^2}{12}$ (\cite{dither}), where $\Delta = 0.5$ px is the quantization increment. However, as the variance of the measured distribution is reduced, this error can depend strongly on the input and is bounded above by $\frac{\Delta^2}{4}$. In future devices, this effect may be reduced through dithering through random temporal or spatial modulation (\cite{dither}).

Another significant source of error is introduced through the spatial calibration procedure. For the present data, a circular target is used spanning $13.6 - 22.1$ px diameter. A conservative estimate of the error from this procedure is $\pm 1$ px in the measured diameter, corresponding to a possible $\pm 4.5 - 7.3 \%$ error in the measured velocity. Although this is likely an overestimate, it is consistent with the errors observed in the benchmark experiments. This procedure could be improved through various super-resolution imaging techniques or by calibrating directly with other velocimetry methods.

In future devices, integrating more advanced or custom sensors may allow for optimized image processing, higher resolution sub-pixel estimation, and general algorithmic control which is tailored for this specific application. 

\section{Statements and declarations}
\subsection{Acknowledgements}
The authors thank Peter Gunnarson for fruitful discussions.
\subsection{Author contributions}
Robert Hunt: Conceptualization, Methodology, Software, Validation, Investigation, Writing – Original draft, Review \& Editing, Funding acquisition. Eli Silver: Methodology, Software, Validation, Resources, Data Curation, Writing – Review \& Editing, Funding acquisition. Daniel M. Harris: Writing – Review \& Editing, Supervision, Project administration, Funding acquisition.
\subsection{Funding}
We gratefully acknowledge funding from the Brown University School of Engineering Hazeltine Innovation Award and the Office of Naval Research (Grant No. N00014-21-1-2670).
\subsection{Competing interests}
The authors have no relevant financial or non-financial interests to disclose.
\subsection{Data availability}
All data presented in this work is provided in the Supplementary Information. Other information can be found at \href{https://github.com/harrislab-brown/FIEVel}{https://github.com/harrislab-brown/FIEVel}.

 \bibliography{sn-bibliography}

 \end{document}